# Teaching Survey Research in Software Engineering

Marcos Kalinowski[0000−0003−1445−3425] and
Allysson Allex Araújo[0000−0003−2108−2335] and
Daniel Mendez[0000−0003−0619−6027]

**Abstract** In this chapter, we provide advice on how to effectively teach survey research based on lessons learned from several international teaching experiences on the topic and from conducting large-scale surveys published at various scientific conferences and journals. First, we provide teachers with a potential syllabus for teaching survey research, including learning objectives, lectures, and examples of practical assignments. Thereafter, we provide actionable advice on how to teach the topics related to each learning objective, including survey design, sampling, data collection, statistical and qualitative analysis, threats to validity and reliability, and ethical considerations. The chapter is complemented by online teaching resources, including slides covering an entire course.

## 1 Introduction

Survey research has become an indispensable tool in software engineering, allowing for gathering cross-sectional insights into opinions, experiences, and needs with respect to practices and problems that can help shape the discipline. In this chapter, we take a particular focus on addressing the challenges we experienced when teaching survey research in software engineering, aiming to provide a comprehensive guide that educators can follow and adapt to their specific contexts.

Drawing from several international teaching experiences among the authors, we start by providing teachers with a comprehensive overview of a what we ourselves

Marcos Kalinowski
Pontifical Catholic University of Rio de Janeiro, e-mail: kalinowski@inf.puc-rio.br

Allysson Allex Araújo
Federal University of Cariri, e-mail: allysson.araujo@ufca.edu.br

Daniel Mendez
Blekinge Institute of Technology and fortiss, e-mail: daniel.mendez@bth.se

consider to be a typical syllabus for teaching survey research. This syllabus includes a detailed course description, clearly defined learning objectives, a schedule of lectures to be taught and examples of practical assignments. Please note that we do not claim (nor intent) to provide structure for a universal way of teaching survey research. Rather, we describe a blueprint that we extracted from our own past experiences. Those experiences emerge from both teaching survey research in various settings (MSc level and PhD level) and applying the core concepts as part of large-scale surveys and families of surveys (*e.g.*, [2], [14], [18], [21], [22], [32], [45]). We trust that the reader will know how to adapt the insights given to the particularities of their context and hope to support them in this endeavour with the selection of topics that we believe are important to cover.

The chapter provides advice for proficiently teaching the main topics related to survey research. These topics include the characteristics and purpose of survey research, survey design, sampling, data collection, statistical and qualitative analysis, threats to validity and reliability, and ethical considerations. Furthermore, the chapter is complemented by online teaching resources, which include slides covering all the lectures. Those resources can be found in the online repository that accompanies this edited book: https://zenodo.org/doi/10.5281/zenodo.11544897.

It is noteworthy that, while the chapter offers a comprehensive guide covering the main topics related to survey research in software engineering, it is also crafted with flexibility in mind. We hope that this facilitates the adoption of the content, methodologies, and resources provided by educators to fit their teaching styles and the specific needs of their students.

The remainder of this chapter is organized as follows. Section 2 presents a what we believe to a typical syllabus for teaching survey research. Section 3 provides advice on how to teach the main topics of survey research, with subsections dedicated to each specific learning objective of the syllabus. Section 4 contains the concluding remarks.

## 2 Survey Research Course Syllabus

This section aims to offer teachers a comprehensive overview of the typical structure, content, and expectations of a course on survey research. We begin with the course description, laying the foundation for what to expect throughout the course. Thereafter, we describe the learning objectives, outlining the skills and knowledge that students are expected to acquire. We then present an example of a schedule planned to deliver a structured and effective learning experience, including assignments to reinforce learning and gauge the students' understanding of the course material. Finally, we discuss assessment possibilities and report on resources that we provide for teachers.

**Course Description.** The course provides a comprehensive overview of survey research principles and practices. Students will be taught how to design and evaluate survey instruments, focusing on aligning them with research objectives and relevant theories. The course covers best practices in sampling, data collection, and both statistical and qualitative analysis techniques. A critical course component is teaching students to identify and address potential threats to the validity and reliability of survey research. Furthermore, it encompasses ethical considerations pertinent to the field of survey research.

**Learning Objectives.** The course structure has been developed to address the learning objectives outlined in Table 1. We also indicate the levels of learning associated with each objective, as categorized by the revised version of Bloom's taxonomy [8].

| ID | Learning Objective | Students will be able to ... | Bloom's Taxonomy |
|---|---|---|---|
| LO1 | Understanding the Characteristics and Purposes of Survey Research | ... articulate on the characteristics and purposes of survey research. ... provide survey research application examples. | Remembering & Understanding |
| LO2 | Designing and Evaluating Survey Instruments | ... create survey instruments aligning with specific research objectives and theories. ... critically assess the effectiveness of survey instruments. | Evaluating & Creating |
| LO3 | Mastering Sampling and Data Collection | ... apply best practices in sampling and data collection. ... understand the trade-offs of different sampling and data collection methods. | Understanding & Applying |
| LO4 | Applying Statistical and Qualitative Analysis Methods | ... utilize statistical and qualitative analysis techniques to interpret survey data. | Applying & Analyzing |
| LO5 | Identifying and Addressing Validity and Reliability Threats | ... analyze and address potential threats to the validity and reliability of survey research. | Analyzing & Evaluating |
| LO6 | Understanding Ethical Considerations in Survey Research | ... identify, understand, and apply ethical considerations in survey research. | Understanding & Applying |

**Table 1** Learning Objectives and Bloom's Taxonomy Levels.

**Schedule.** Our experiences in teaching survey research range from brief and informative formats, like 2-hour seminars or 4-hour conference tutorials, to more comprehensive, in-depth treatments as part of doctorate-level programs. Given the importance of the subject and its evolving nature in software engineering, a more extensive format that includes practical assignments and detailed discussions for each learning objective is advisable. For illustrative purposes, we present the schedule of a recent experience of the first author teaching an 18-hour doctorate-level discipline on the topic at *Universidad de Castilla-La Mancha* (UCLM) in Spain in Table 2. This particular discipline was structured into 12 classes, each lasting 1.5 hours.

It is important to note that this 18-hour schedule assumes that students will independently acquire more detailed knowledge on certain topics, such as statistical and qualitative analysis, outside the course's scope. Without this prerequisite, the course could extend far beyond 18 hours, including supplementary exercises and hands-on practice for statistical and qualitative analyses based on open survey datasets.

| Class | LO | Lecture | Assignment |
|---|---|---|---|
| 1 | LO1 | Characteristics and Purposes of Survey Research | Provide three application examples of survey research in software engineering. |
| 2 | LO2 | Survey Design: Basics of Survey Design | Design basic demographic questions to characterize software professionals. |
| 3 | LO2 | Survey Design: Goal-Question-Metric-Based Design | Break down examples of GQM goal definitions into questions, metrics, and survey questions. |
| 4 | LO2 | Survey Design: Theory-Driven Design | Discuss example papers using theory-driven survey design, identifying hypotheses, underlying theory, constructs, and validated scales. |
| 5 | LO2 | Survey Design: Survey Instrument Evaluation | Discuss survey instrument evaluation methods. |
| 6 | LO3 | Sampling and Data Collection | Provide examples for target populations, units of analysis, sampling strategies, invitation formats, and benchmarks for representativeness. |
| 7 | LO4 | Statistical Analysis: Descriptive Statistics | Apply descriptive statistics on demographic questions from a collected sample using open survey data and argue for sample representativeness. |
| 8 | LO4 | Statistical Analysis: Inferential Statistics | Discuss examples of published surveys applying inferential statistics, including hypothesis testing, bootstrapping with confidence intervals, Bayesian analysis, and structural equation modeling. |
| 9 | LO4 | Qualitative Analysis | Discuss examples of published surveys applying grounded theory procedures and thematic analysis. |
| 10 | LO5 | Threats to Validity and Reliability | Discuss examples of threats to validity and reliability and mitigation actions of published surveys. |
| 11 | LO6 | Ethical Considerations | Draft an informed consent form and discuss ethical concerns and how to report ethical considerations properly. |
| 12 | LO1 - LO6 | Final Survey Protocol Presentations | Present the final survey protocol. |

**Table 2** Course Lectures and Assignments.

**Assignments.** Considering the example course schedule, students are expected to participate actively in each session's assignments. These assignments should ideally be prepared in interactive formats to foster group discussions, e.g., using previously prepared collaborative whiteboard platforms such as Miro[1] or Mural[2]. Furthermore, besides participating in the sessions, throughout the course, students are expected to apply what they have learned to develop a survey protocol for a topic of their choice to be delivered at the end of the course. This protocol should include details on the survey goal, survey design - linked to the underlying theory and related work, intended instrument evaluation, sampling and data collection strategies, statistical and qualitative analysis methods to be applied, threats to validity and mitigation actions, and ethical considerations. Depending on the number of students, this might have to be a group assignment, as students are expected to present an overview of their protocols at the end of the course.

**Assessment.** The assessment method can certainly be adapted by the teacher. However, we recommend allocating 50% of the final grade to participation in the class assignments and the remaining 50% to the quality of the final survey protocol delivered by the students.

## 3 Advice on How to Teach Survey Research

We organize the advice on important aspects to cover when teaching lectures related to survey research based on the learning objectives detailed in the previous section. The recommendations are based on lessons learned from teaching and from experiences conducting large-scale international surveys.

### 3.1 Teaching Characteristics and Purpose of Survey Research (LO1)

In our example schedule (*cf.* Table 2), this learning objective is taught in a single lecture. Hereafter, we provide some recommendations for teaching the characteristics and purposes during this lecture.

#### 3.1.1 Characteristics of Survey Research

When teaching the characteristics of survey research, it is important to cover the following aspects to ensure students grasp the potential but also the limitations of this method.

**Prevalence of Surveys.** Highlight how surveys are widely used across various fields, emphasizing their utility in capturing information from large groups of individuals [15, 46]. It is advisable to cite examples of relevant software engineering surveys and their findings to motivate the students. Some well-known examples follow. In Brazil, the *iMPS* survey, conducted six years in a row with hundreds of software organizations [23, 44], allowed the assessment of the impact of adopting a software process reference model on performance results [22]. The *Naming the Pain in Requirements Engineering (NaPiRE)* family of surveys revealed the requirements engineering status quo [45] and contemporary problems [14] in conventional and in machine-learning contexts [2, 21]. The *HELENA* survey transferred the principles underlying the NaPiRE survey to provide evidence on the prevalence of hybrid development approaches and allowed a reinterpretation of agility twenty years after the agile manifesto [25].The *Pandemic Programming* survey revealed how COVID-19 affected the well-being and productivity of software developers [32]. Surveys have been particularly useful in assessing human aspects related to software engineering. For instance, the survey by Graziotin *et al.* [17] assessed the consequences of happiness and unhappiness among software developers. Another international survey on the impostor phenomenon revealed a high prevalence of impostor feelings among software engineers, especially women [18]. Providing examples of relevant scientific findings will help to motivate your students.

---

[1] http://www.miro.com/

[2] http://mural.co/

**Timing of Surveys.** Discuss the timing of survey administration relative to the occurrence of the phenomena of interest. Explain how surveys can be retrospective (looking back at something that has already happened) or prospective (looking ahead to something that is expected to happen). Remind the students that in a survey, they can either ask for specific facts that participants experienced or for opinions on topics, but that, in general, from a scientific perspective, facts are preferred over opinions [42].

**Lack of Experimental Control.** Clarify that, unlike controlled experiments, surveys do not allow for control over variables or direct manipulation of the environment. Emphasize the observational nature of survey research, which often leads to challenges in establishing causality.

**Focus on Understanding.** Emphasize the importance of designing surveys to maximize understanding from a minimal set of variables. Teach students about the trade-offs between the breadth of information collected, the depth of understanding, and the importance of focused research questions.

**Generalization to the Population.** Ensure that students understand that surveys are rarely conducted to create an understanding concerning a particular sample; the typical focus is on generalizing results to the population from which the sample was drawn. Surveys that conclude only on a given sample based on descriptive statistics, without applying inferential statistics for generalization purposes (or at least generalization by analogy in relation to existing evidence), are, from a scientific point of view, not as powerful as they could be for helping to building and evaluating theories, i.e. they would barely contribute to scientific progress.

### 3.1.2 Purpose of Survey Research

When teaching the purposes of survey research, it is essential to cover the distinct general objectives that surveys can fulfill [46, 47]. We provide examples to explain these objectives precisely.

**Exploratory Surveys.** Emphasize the exploratory nature of these surveys as a preliminary step in research. Discuss how exploratory surveys can help identify key issues and constructs that should not be overlooked in subsequent in-depth investigations (*e.g.*, constructs of a theory, like requirements elicitation techniques [45]). Provide examples of how these surveys can uncover new areas of interest, such as emerging techniques in a field like requirements elicitation. Such surveys may also serve the purpose of preparing for a larger investigation where the identified constructs can further be investigated by complementary research methods.

**Descriptive Surveys.** Explain how descriptive surveys depict the characteristics of a population or phenomena. Teach the importance of clear and focused survey questions to accurately capture things like the distribution of attributes within a population (*e.g.*, usage of requirements elicitation techniques [45]). Remember to mention the importance of analyzing and interpreting the collected data beyond the sample to make assertions about the population (*e.g.*, using inferential statistics), such as the prevalence of certain practices.

**Explanatory Surveys.** Define explanatory surveys and their goal to establish cause-and-effect relationships. Discuss how these surveys can provide insights into why certain phenomena occur, such as the reasons behind using specific techniques in certain contexts. For example, why are specific requirements elicitation techniques used in specific contexts [45]. Highlight the role of designing surveys with questions that can pave the way to explanatory findings, coupled with the proper utilization of statistical techniques for data analysis; this will help to motivate students for these future learning topics.

### 3.1.3 Assignments

After teaching the characteristics and purposes of survey research, a recommended assignment involves conducting a brainstorming session, asking each student to think out of the box about three application examples of survey research in software engineering. A collaborative whiteboard platform, such as Miro or Mural, could be used for this purpose, and the individual contributions of the students could be collectively grouped by affinity by the instructor. This assignment could also break the ice and organize the students into topic affinity groups to work together on the final survey protocol throughout the course.

> **Key Takeaways on Teaching Characteristics and Purposes**
>
> - Precisely explain the characteristics of survey research to ensure students understand the strengths and limitations of the method.
> - Cover the distinct general objectives that surveys can fulfill, providing examples to ensure students understand the method's investigation possibilities.

### 3.2 Teaching Designing and Evaluating Survey Instruments (LO2)

As outlined in our example schedule (*cf.* Table 2), this learning objective typically requires several lectures. We recommend at least one lecture on the basics of survey design, two additional ones on specific and effective design approaches that we have been teaching, and a last one on survey instrument evaluation methods. We abstracted and conceptualized the two specific design approaches based on best practices observed in several surveys published in top-tier software engineering journals and herein introduce them to the community as *Goal-Question-Metric-Driven (GQM-Driven) Survey Design* and *Theory-Driven Survey Design*. Recommendations for these lectures follow.

#### 3.2.1 Basics of Survey Design

This lecture is a prerequisite for the subsequent ones, which cover specific survey design approaches. When teaching the basics of survey design, it is important to provide a structured approach that encompasses various types of questionnaires, question types, and question categories, as well as measurement scales and conditions for obtaining accurate responses. Incorporating these elements into teaching will help students understand the fundamental principles of survey design and develop the skills necessary to create effective and reliable survey instruments. Recommendations for teaching them follow.

**Questionnaire Types.** Explain the difference between self-administered and interviewer-administered questionnaires, including their advantages and contexts of use. Discuss the importance of considering the respondent's environment and potential distractions for self-administered questionnaires. For interviewer-administered questionnaires, highlight interviewer effects and the need for standardization in question delivery.

**Question Types.** Demonstrate how to construct open-ended, close-ended, and hybrid questions, including when and why to use each type. Practice writing questions with the class and analyze examples for clarity and bias.

**Question Categories.** Define demographic, substantive, filter, and sensitive questions, providing advice on how to integrate them into a survey. Discuss ethical considerations when asking sensitive questions and the use of filter questions to guide respondents through the survey.

**Measurement Scales.** Teach the characteristics of nominal, ordinal, interval, and ratio scales and provide examples of each. Use exercises to help students choose the appropriate scale for different types of data they want to collect.

**Conditions for Appropriate Responses.** Emphasize the need for questions to be understandable, ensuring they match the target population's language proficiency and cultural context. Ensure that respondents of that population (and your sample) have the knowledge to answer the questions accurately and discuss strategies to encourage their motivation and willingness to participate.

**Common Question Wording Problems.** Offer strategies to avoid common question-writing pitfalls such as using complex language, technical terms, double-barreled questions, and double negatives. Highlight the importance of keeping questions concise to prevent misunderstanding and respondent fatigue.

**Opinions vs. Facts.** Differentiate between questions that seek subjective opinions and those that ask for more factual phenomena (observed or experiencd by the respondents). Discuss the relevance of each in relation to the survey's goals and the implications for data analysis.

**Assignments.** A recommended assignment after teaching the basics of survey design involves asking students to design basic demographic questions, for instance, to characterize software professionals. Engage students to review their questions critically and to compare them to questions contained in well-known surveys that also involve software professionals, such as the Stack Overflow Annual Developer survey [31]. Explain to them, however, that even the demographics to be collected strongly depend on the goal of the survey. Better aligning survey questions to research goals is the topic of the next suggested lecture.

### 3.2.2 Goal-Question-Metric-Driven Survey Design

This lecture focuses on teaching a specific survey design approach that helps to design concise surveys focused on relevant goals and research questions. Although we are conceptualizing it as part of this book chapter, we have been teaching it for years, and it has been explicitly or implicitly applied in several software engineering surveys that carefully describe the rationale of deriving questions and metrics from a clear research goal (*e.g.*, [2, 7, 18, 22, 23, 25, 29]). Advice on how to teach the approach and assignments follows.

The *Goal-Question-Metric-Driven Survey Design* approach adapts the more generic Goal-Question-Metric (GQM) measurement approach [5, 6] to survey design. GQM defines a generic way to plan and execute measurement and analysis activities. It starts with the declaration of the measurement *Goals*. From the goals, *Questions* that we would like to answer with the data interpretation are defined. Finally, from the questions, the *Metrics* and the data to be collected are defined. The adaptation for survey design consists of using the GQM goal definition template to precisely describe the purpose of the survey, generating research questions to be answered from the goal definition, and then generating metrics collected to answer these questions. The survey instrument is then designed following the basic survey design principles with questions to collect these metrics objectively.

We recommend teaching the GQM-driven survey design approach by providing examples of precise goals defined by applying the GQM goal definition template to surveys. The GQM goal definition template is defined as follows.

```
Analyze <object of study>
with the purpose of <goal>
with respect to <quality focus>
from the point of view of the <perspective>
in the context of <context>.
```

We recommend to pay special attention when teaching to fill this template for surveys correctly. The object of study corresponds to the phenomena of interest to be investigated. The goal is typically characterizing (for explorative or descriptive surveys) or understanding (for explanatory surveys). The quality focus will be directly related to the metrics to be collected through the survey's substantive questions. The perspective should reflect the survey population (in the case of more direct questions) or the researcher (in the case of using more robust validated scales to be interpreted by the researcher). Finally, the context allows further detailing of the data collection context.

An example goal definition, taken from the iMPS survey [23], which assessed the performance of software organizations that adopted the Brazilian software process reference model (MPS-SW [35]), follows. "Analyze *the profile of software development organizations* with the purpose of *characterizing* with respect to *the organizations' current profile, satisfaction degree regarding the MPS model, variation of presence in international markets, variation of exportation volume, and variation concerning cost, estimation accuracy, productivity, quality, user satisfaction, and return of investment (ROI)* from the point of view of *software development organizations* in the context of *software development organizations with unexpired MPS-SW assessments published in the SOFTEX portal*."

Following the GQM-driven survey design approach, this goal was detailed into questions, which were then detailed into metrics [44]. The survey was then carefully designed to capture these metrics. To exemplify this process, estimation accuracy is one of the elements of the quality focus. To capture this quality focus element, the question was "What is the organizations' estimation accuracy?". However, this is a research question and not yet a survey question. First, we have to break it down into metrics. The metrics that we were looking for were the *average project duration* and the *average estimated project duration*. Based on these base metrics, we could then calculate the estimation accuracy. Hence, the estimation accuracy element led us to design two survey questions: "What is the average duration of projects conducted within the last 12 months, measured in months?" and "What is the average estimated duration of projects conducted within the last 12 months, measured in months?".

This example leads us to an important aspect that should be emphasized when teaching this approach. The GQM questions do not necessarily directly relate to the survey questions. Instead, the survey questions should be designed to focus on the metrics (and scales) and the data to be collected. There are several examples of published surveys explicitly using the GQM goal definition template and the question and metrics breakdown rationale to support designing the survey (e.g., [7, 12, 18, 29]).

**Assignments.** A recommended assignment after teaching GQM-driven survey design involves providing students with examples of GQM goal definitions taken from research papers, asking students to break them down into questions and metrics. Thereafter, they should be asked to design survey questions to collect these metrics from population sample. Using the GQM goal definition template from the survey by Guenes *et al.*, which concerns the impostor phenomenon, could allow an interesting reflection linking to the next lecture, as it will break down to metrics that correspond to theoretical constructs (*e.g.*, impostor phenomena score or perceived productivity) that will require more complex approaches, such as validated scales and careful theoretical construct breakdowns, for proper measurement.

### 3.2.3 Theory-Driven Survey Design

This lecture focuses on teaching a more robust survey design approach that helps to design surveys to support theory building and evaluation. A theory provides explanations and understanding in terms of basic constructs and underlying mechanisms. From a practical perspective, theories are useful and explain or predict phenomena that occur in software engineering; from a scientific perspective, theories should guide and support further research in software engineering [20].

We conjecture that the type of research to be supported by theories includes survey research. In fact, theory building and evaluation, as elaborated in the introductory editorial chapter to the edited book this chapter is part of, can guide the design and analysis of surveys, and surveys can also be applied to evaluate theories [46]. Although we are conceptualizing the theory-driven survey design approach as part of this chapter, we have been teaching it for years; theory has guided the design of several robust software engineering surveys (e.g., [18, 17, 32, 45]). Advice on how to teach the approach and assignments follows.

Theory building and survey research can be strongly interrelated. The *Theory-Driven Survey Design* approach relies on drafting a theory based on observations and available literature and designing the survey based on that theory to test or potentially extend it. According to Sjoberg *et al.* [37], the main building blocks of such a theory are constructs, propositions, explanations, and scope. Constructs describe what the basic elements are, propositions how the constructs interact with each other, explanations why the propositions are as specified, and the scope elaborates what the universe of discourse is in which the theory is applicable.

An initial theory may be as simple as a taxonomy of constructs or a set of statements relating to constructs. For the NaPiRE [45] and the Impostor Phenomenon [18] surveys, a set of constructs and propositions (hypotheses to be tested) were elaborated based on available literature and expert knowledge. For the Pandemic Programming survey [32], a multifactorial theoretical model relating a set of constructs was designed based on related work. In both cases, the surveys were designed to test the theory using inferential statistics. We will use the theoretical model of the pandemic programming survey to highlight some key issues to be emphasized when teaching theory-driven survey design.

The theoretical model of developer well-being and productivity, which supported the design of the Pandemic Programming survey, is shown in Figure 1. The survey assessed ten hypotheses revolving around the five constructs shown in the model. H1 and H2 stated that working from home due to COVID-19 would reduce well-being and perceived productivity, respectively. The arrows depict the remaining eight hypotheses, and more details can be obtained in the paper [32]. It is important to highlight that the constructs and the hypotheses should not be randomly drawn. Someone applying theory-driven survey design should do their homework, looking for related work in the literature to identify theoretical constructs and hypothesize meaningful relations, explaining the rationale behind such relations. For instance, the rationale for H1 is based on related work indicating that uncertainty and isolation induce feelings of frustration, anxiety, and fear [11, 39, 40]. Therefore, it would be likely for developers to experience reduced emotional well-being.

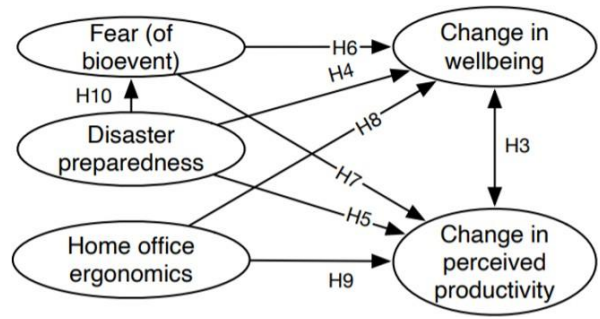

**Fig. 1** Exemplary theoretical model of the developer well-being and productivity survey [32].

Another relevant aspect of the theory-driven survey design approach that should be highlighted when teaching the subject is to use validated scales as part of the survey to measure the constructs, as much as possible, to improve the survey's construct validity. For instance, the pandemic programming survey used the World Health Organization's five-item well-being index (WHO-5) to assess emotional well-being. The WHO-5 scale is widely used, widely applicable, and has high sensitivity and construct validity [41]. Similarly, the Impostor Phenomenon survey used the Clance Impostor Phenomenon Scale (CIPS) to measure how often and seriously the impostor phenomenon interferes with a person's life. CIPS is internationally validated and is the most used scale by researchers and practitioners [27].

In summary, survey research and theory building are strongly interrelated. The exact relationship depends on whether the theory is descriptive, explanatory, or predictive. Hence, theories are of high value to guide the design of robust surveys. On the other hand, survey data supports the definition or refinement of constructs, relationships, explanations, and the scope of a theory, as well as the testing of a theory. Validated scales should be used as much as possible to improve construct validity. So far, the software engineering community still falls short in maturity concerning building validated scales, which is an inherently complex task that has, however, been successfully achieved in other fields, such as psychology [13]. Building (and using) robust scales could help to take the maturity of survey research in software engineering to the next level. These discussions should be part of this lecture.

**Assignments.** A recommended assignment after teaching Theory-driven survey design involves providing students with examples of research papers using the theory-driven approach (e.g., [18, 32, 45]), asking the students to identify theoretical constructs, related scales, hypotheses (or propositions), and underlying theory. As part of our complimentary material, we provide a Miro board where students can collectively work on a research paper registering the demographics and related benchmarks, the constructs and related scales, and the hypotheses (or propositions) and their related underlying theories.

### 3.2.4 Survey Instrument Evaluation

This final lecture on survey design focuses on teaching instrument evaluation methods. Several survey instrument evaluation methods can be used to assess the validity and reliability of the survey instrument. We recommend covering the following methods, which are typically related to this purpose [26, 34]: expert reviews, focus groups, pilot surveys, cognitive interviews, and experiments. Additionally, the empirically evaluated checklist for surveys in software engineering by Molléri *et al.* [30] can be used as an additional valuable resource for evaluating the survey design (as well as the final survey report).

Encourage students to understand the role of expert reviews in assessing content validity. Discuss how to select experts and what criteria they should use to evaluate the survey questions. With respect to focus groups, explain how they can be utilized to gather feedback on the survey's comprehensibility, relevance, and overall design. Highlight the importance of diverse participant selection to ensure a wide range of perspectives. Pilot surveys are highly recommended. Teach the purpose and process of conducting pilot surveys, emphasizing their role in identifying issues with survey instructions, question-wording, and format. Discuss how to analyze pilot survey data to make necessary adjustments before full deployment. Remember to describe cog-nitive interviewing techniques for evaluating how respondents perceive and process survey questions. You may practice mock cognitive interviews in class to give students a hands-on experience. Finally, introduce the concept of using experimental designs to test different aspects of the survey instrument, such as question order effects. Discuss how to set up control and experimental groups and measure the impact of changes to the survey.

**Assignments.** The assignment for this lecture may involve discussing the different survey instrument evaluation methods and how they were applied in surveys published in the software engineering realm.

> **Key Takeaways on Teaching Designing and Evaluating Survey Instruments**
>
> - Do not forget the basics! Teach the "when" and "why" of different types of questionnaires, question types, and question categories, as well as measurement scales and conditions for obtaining accurate responses.
> - Teach GQM-Driven and Theory-Driven survey design approaches to help students design focused and concise surveys, maximizing understanding based on a minimal set of variables related to relevant research goals or theories.
> - Emphasize the importance of using validated scales as much as possible to improve construct validity.
> - Emphasize that survey instruments should be evaluated using different methods to avoid threats to validity and improve reliability.

### 3.3 Teaching Sampling and Data Collection (LO3)

The third learning objective covers sampling and data collection. In our example schedule (*cf.* Table 2), it is taught in a single lecture. However, when having an entire course dedicated to survey research, there would surely be space to devote more attention to this important topic. Hereafter, we provide some recommendations during this lecture when teaching each of these topics.

#### 3.3.1 Sampling

Baltes and Ralph [4] critically reviewed sampling in software engineering research and found that sampling, representativeness, and randomness often appear misunderstood, highlighting the importance of properly teaching these aspects. Hence, when teaching survey research sampling, it is critical to define the target population precisely and impart the concepts of representativeness and sample size. Different sampling strategies and their intricacies should also be taught. We provide recommendations on these elements hereafter.

**Defining the Target Population.** Start by discussing the importance of clearly identifying the target population the survey aims to generalize to. Use real-world examples to illustrate how the unit of analysis can vary in software engineering surveys [28], such as organizations, software teams or projects, and individuals. Stress the importance of clearly defining the target population to ensure the survey's relevance and applicability (and reproducibility).

**Representativeness.** Explain the concept of representativeness and its significance in generalizing survey results to the broader population. Highlight the challenges in software engineering research due to the lack of comprehensive demographic information from traditional sources like government statistical offices. Discuss alternative sources of demographic data, such as commercial data providers and community-driven surveys like the Stack Overflow Annual Developer Survey [31]. They have the bias that only people registered at Stack Overflow can be sampled. Yet, this could be tolerated in light of the popularity of the platform among software developers. If possible, conduct a class activity to analyze the demographic data from these sources and discuss how they can help to define the sampling strategy.

Teach students how to compare the demographics of their survey sample with larger datasets to assess representativeness. Emphasize the importance of transparency in reporting the degree of representativeness to prevent overclaiming and to enhance research credibility. An example of a survey with such a representativeness assessment visually comparing the demographics of the gathered sample against those of the Stack Overflow Annual Developer Survey can be seen in the Impostor Phenomenon survey [18].

**Sample Size Estimation.** Introduce the concept of sample size and its critical role in statistical analysis and the conclusion validity of surveys. Explain formulas for calculating appropriate sample sizes for a population. A simple way, for example, is to follow Yamane [48]. As described by Wagner *et al.* [46], he proposed to use the following equation to calculate a suitable sample size $n$:

$$n = \frac{N}{1 + Ne^2} \tag{1}$$

In the Equation 1, $N$ is the population size and $e$ is the level of precision. For the population estimate of 23 million developers worldwide and a level of precision of 0.05, this would require a sample of size 400. Hence, explain that for most intents and purposes, a sample size of more than 400 allows to claim strong generalizability given that the representativeness was also checked, as previously described. To avoid raising expectations, inform that most surveys may fall short of this, but explain that a clear discussion comparing the sample size and representativeness with these figures makes it easy to evaluate the strengths and weaknesses of a particular survey.

**Sampling Strategies.** Survey sampling strategies are crucial to understand because they directly impact the validity and generalizability of survey research results. Linaker *et al.* [26] present some common sampling strategies, dividing them into non-probabilistic and probabilistic sampling. Hereafter, we briefly introduce sampling strategies to be covered, along with some advice for teaching them.

- **Non-probabilistic sampling**. Related to all sampling approaches in which randomness could not be observed in selecting the units.

  - *Convenience (Accidental) Sampling*. Choosing individuals who are easily accessible or willing to participate, such as students in a classroom. Discuss the limitations of convenience sampling, such as potential bias.
  - *Quota Sampling*. Ensuring that the sample includes certain percentages of groups, reflecting their proportions in the target population. Create scenarios for students to design quota samples based on given population characteristics.
  - *Purposive (Judgement) Sampling*. Selecting participants based on specific characteristics or qualities, making them suitable for answering the research question. Engage students in discussions about when purposive sampling is appropriate (e.g., expert surveys, such as [29]) and have them outline a plan for selecting a purposive sample.
  - *Snowball Sampling*. Asking participants to refer others to the study, which is particularly useful for hard-to-reach or specialized populations. Discuss the potential implications of participant referrals.

- **Probabilistic sampling**. All units from a sampling frame must have the same probability of being selected.

  - *Simple Random Sampling*. Selecting individuals from a larger population where each individual has an equal probability of being chosen. Demonstrate the use of random number generators or other tools to achieve a random sample.
  - *Clustered Sampling*. Dividing the population into clusters (e.g., geographic areas) and randomly selecting within these clusters. Provide cases where cluster sampling is appropriate.
  - *Stratified Sampling*. Dividing the population into strata (e.g., age, income level) and then randomly sampling from each stratum to ensure representation across key variables. Teach how to stratify your sample based on distributions of the populations to increase sample representativeness.
  - *Systematic Sampling*. Choosing every nth individual from a list after a random start. Discuss that, in general, systematic random sampling is preferred but that the systematic sampling approach can also have benefits, such as assuring that the sample is spread evenly over a population, reducing the risk of accidental clustering.

### 3.3.2 Data Collection

There are essentially two strategies to approach the target population to collect data in questionnaire-based surveys, having very distinct implications on the survey design and the recruitment approaches [46]:

- *Closed invitations* follow the strategy of approaching known groups or individuals to participate in a survey per invitation only and restrict the survey access only to those being invited.
- *Open invitations* follow the strategy of approaching a broader, often anonymous audience via open survey access; i.e., anyone with a link to the survey can participate.

When teaching data collection in survey research, the implications of these strategies should be precisely explained to enable informed decisions. We hereafter summarize implications that should be taught, based on challenges reported by Wagner *et al.* [46]. Closed invitations are suitable when it is possible to precisely identify and approach a well-defined sample of the target population. They may be required when filtering out participants who are not part of the target population would be difficult, harming the sample representativeness. Open invitations, on the other hand, allow reaching out for larger samples. However, they typically require more careful consideration of context factors when designing the survey instruments. These context factors can then be used during the analyses to filter out participants that are not representative (*e.g.*, applying the blocking principle to specific context factors), improving the survey instrument's criterion validity.

### 3.3.3 Assignments

As survey data collection is hard to enact in teaching scenarios, we recommend an assignment for the students to provide and discuss examples of target populations, detailing the units of analysis, sampling strategies, invitation formats, and benchmarks for representativeness.

> **Key Takeaways on Teaching Sampling and Data Collection**
>
> - Teach sampling fundamentals and strategies (and their implications). Sampling, representativeness, and randomness are often misunderstood in software engineering research.
> - Discuss the strategies to approach the target population and their implications to allow students to make informed decisions when planning their surveys.

## 3.4 Teaching Statistical and Qualitative Analysis (LO4)

The next learning objective covers statistical and qualitative analysis. We recommend lectures on descriptive and inferential statistics and an additional lecture on qualitative analysis. Of course, the depth in which these lectures are covered will depend on the students' background. Recommendations for these lectures follow.

### 3.4.1 Statistical Analysis

Although surveys can be qualitative, most often, the majority of the questionnaires are composed of closed questions that have quantitative results. Therefore, it is fundamental to teach the main concepts of statistical analysis and show the students their possibilities for analyzing quantitative survey results. In particular, we recommend emphasizing that the statistical analysis should compromise the sample's descriptive statistics and inferential statistics to generalize the results to the population.

**Descriptive Statistics.** The goal of descriptive statistics is to characterize the answers to one or more questions of our specific sample. In surveys, they provide a way to summarize large samples to make them understandable at a glance. Items to be covered when teaching descriptive statistics include measures of central tendency, measures of variability, frequency distributions, and graphical representations.

Which descriptive statistic is suitable depends on what we are interested in most and the scale of the data. Teach how to choose the appropriate measures and graphical representation depending on the data type and what you want to convey.

For ordinal scales (*e.g.*, Likert scales), we can use frequency counting, mode, median, minimum, maximum, median absolute deviation, and interquartile range. For graphical representation, we recommend providing examples showing the distribution of ordinal data in stacked bar charts.

For interval or ratio scales, we can use all available descriptive statistics, such as mean, variance, and standard deviation. For graphical representation, we recommend providing examples using boxplots, which provide a descriptive overview and enable eliminating outliers by using the quartile method.

**Inferential Statistics.** While descriptive statistics concern the sample, inferential statistics allow making predictions and generalizations about a population. Typically these statistics are the ones used to answer the research questions defined in the survey design. When teaching inferential statistics, we recommend explaining different possibilities for analyzing quantitative survey results, including null hypothesis significance testing, bootstrapping with confidence intervals, bayesian analysis, and structural equation modeling.

As the name suggests, null hypothesis significance testing is particularly useful for testing hypotheses related to a theory. Propositions can be operationalized into hypotheses to test with the survey data. In surveys, we typically can have point estimate hypotheses for answers to a single question or hypotheses on correlations between answers to two questions. Several statistical significance tests can be applied, with differences in their statistical power. Of course, these tests require a representative sample. Examples of such tests include Mann-Whitney, t-test, Wilcoxon signed-rank, paired t-test, Kruskal-Wallis, ANOVA, and Tukey. The characteristics of such tests and their assumptions should be taught. Ideally, the statistical test for which the assumptions are met with the highest power should be used to evaluate the hypotheses.

One of the limitations of null hypothesis significance testing lies in its dichotomous nature. An alternative lies in using bootstrapping to calculate confidence intervals [46], which allows for avoiding fixed significance level thresholds. The idea of bootstrapping in this context is to repeatedly take samples with replacements from the survey sample and calculate the statistic we are interested in. This is repeated a large number of times and, thereby, provides us with an understanding of the distribution of the sample. An example taken from the NaPiRE survey [45] is shown in Figure 2, where the confidence intervals allow observing that more than 60% of the organizations elicit requirements using interviews or facilitated meetings.

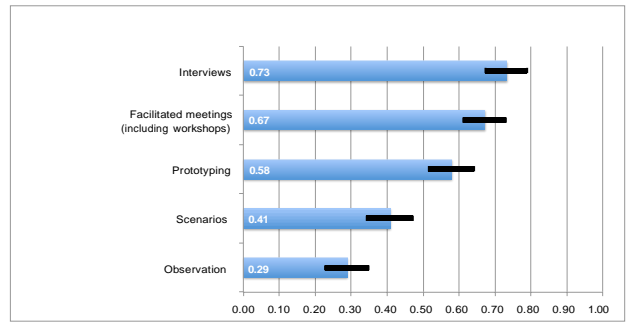

**Fig. 2** Proportions with confidence intervals for question "How do you elicit requirements?" [45].

Another alternative comprises Bayesian analysis [43], where probability is understood as a representation of the state of knowledge or belief. This approach acknowledges uncertainty and allows the gradual integration of evidence and accumulating knowledge. It is suitable for situations in which you are building knowledge based on several survey trials.

Finally, structural equation modeling can be employed to analyze multiple variables at the same time, allowing an understanding of complex relationships between variables to validate theoretical models. The *Pandemic Programming* [32] survey represents an application example of this statistical analysis technique in the context of software engineering surveys.

#### 3.4.2 Qualitative Analysis

When teaching qualitative analysis in the context of survey research, emphasize the role of open questions. Open questions allow respondents to freely express their perspectives and perceptions, providing a rich dataset for analysis. Highlight that these types of questions can provide detailed insights into a phenomenon, potentially revealing causal relationships among theoretical constructs and leading to new theoretical explanations. However, clarify that analyzing the resultant large volumes of qualitative data is not trivial and requires substantial effort.

We recommend introducing Thematic Analysis [9] and Grounded Theory [16] as pivotal methods for supporting qualitative analyses of data derived from open questions. Thematic Analysis allows for the identification, analysis, and reporting of themes within data, offering a flexible approach that can be either deductive or inductive, depending on the research goals. Grounded Theory, on the other hand, involves inductively generating theories from data, helping researchers to systematically identify themes and build theoretical frameworks from the patterns that emerge in the qualitative responses.

It is important to emphasize that correctly applying qualitative methods requires a deep understanding of their respective procedures and nuances. For instance, grounded theory has several variants, and Stol *et al.* [38] provide specific considerations for properly conducting and reporting grounded theory. For further advice on teaching qualitative methods, we recommend referring to Christoph Treude's chapter on Qualitative Data Analysis in Software Engineering by Christoph Treude in this edited book our chapter is part of.

### 3.4.3 Assignments

Recommended assignments related to teaching statistical and qualitative analysis organized based on the class schedule in Table 2 include the following:

**Descriptive statistics assignment:** Students access an open survey dataset, perform descriptive statistics on demographic data, and assess the sample's representativeness by comparing these statistics to established benchmarks. Guenes *et al.* [18] provide an example survey providing such an open dataset and including representativeness analysis.

**Inferential statistics assignment:** Students review survey research articles that utilize different inferential statistic methods, such as hypothesis testing (*e.g.*, [10], [18], [45]), bootstrapping (*e.g.*, [2], [18], [45]), and structural equation modeling (*e.g.*, [19], [32]). Each student should select one article, summarize the use of inferential statistics, and critically assess how these methods helped address the research question.

**Qualitative analysis assignment:** Students review survey research articles that involve qualitative analysis and apply procedures from grounded theory and/or thematic analysis (e.g., [14], [21]). For each article, students will describe how the method was used, assess its effectiveness in contributing to the research goal, and identify any limitations.

It is noteworthy that, focusing on the specifics of survey research, we did not include more in-depth hands-on inferential or qualitative analysis assignments. This may, however, vary according to the specific learning objectives.

> **Key Takeaways on Teaching Statistical and Qualitative Analysis**
> - Teach descriptive statistics and alternatives for inferential statistics, showing the students their possibilities for analyzing quantitative survey results.
> - Emphasize the role of open questions and teach alternatives for qualitative analysis, with explanations of their adoption rationales.

### 3.5 Teaching Threats to Validity and Reliability (LO5)

Validity is a property of inferences, and the focus of surveys is on reaching conclusions that involve making inferences about the population. Hence, every survey faces threats to validity. As outlined in our example schedule (*cf.* Table 2), a lecture dedicated to this topic could address this learning objective.

The notions of validity and reliability are central to understanding the thoroughness and trustworthiness of the survey [24] and should be taught with special attention. Validity refers to whether the questionnaire measures what it is supposed to. Reliability regards if the results are generalizable. Hereafter we provide advice on how to teach about validity and reliability in survey research.

#### 3.5.1 Validity

In psychometrics, the validity of a survey instrument (test) concerns "the degree to which evidence and theory support the interpretation of test scores for proposed uses of tests" [1]. However, we mainly lack such mature instruments in software engineering. While we should rely on validated instruments as much as possible, software engineering survey instruments are typically designed in a customized way based on specific research goals and explorative, descriptive, or explanatory purposes. Therefore, we typically aim to assess validity more informally, discussing whether it is possible to safely conclude that a survey measures what it is supposed to measure. We recommend teaching and discussing the following four types of validity for software engineering survey instruments:

- *Face Validity.* Focuses on the appearance of the survey and its instrument and refers to how suitable the content of a test appears to be on the surface. Mitigating face validity threats typically involves a lightweight review of the questionnaire by randomly chosen respondents.
- *Content Validity.* Concerns comprehensive coverage of the subject matter. Mitigating content validity threats typically involves a group of reviewers evaluating the questionnaire. The group should include subject matter experts to ensure accurate usage of content-related concepts and respondents from the target population to ensure that the terms used for these concepts are easily understandable to that population.
- *Criterion Validity.* Measures how effectively a survey predicts outcomes or correlates with other established benchmarks. It indicates if a survey works well in practical situations. Mitigating criterion validity issues in software engineering surveys typically involves assuring that the survey can accurately separate respondents from the target population and comparing the results to establish benchmarks for that population.
- *Construct Validity.* Concerns theoretical soundness of the survey. It assesses how well the questions actually measure the constructs they were intended to measure. A way to mitigate construct validity concerns is to rely on validated instruments for the constructs involved in the survey as much as possible and to clearly include the rationale behind any needed adaptation.

### 3.5.2 Reliability

Teaching reliability (aka external validity or generalizability) in survey research involves educating students on how to ensure that their surveys produce dependable results that can be properly generalized for the population [33].

We recommend teaching and discussing reliability concerns that are relevant to survey research in software engineering, which include:

- *Internal Consistency.* Measures whether different items on the same survey produce consistent results. In software engineering surveys, this could involve triangulating answers to related questions (*e.g.*, productivity, size, and effort), or even require adding some extra questions for validation purposes.
- *Test-Retest Reliability.* Assesses the consistency of results when the same survey is administered to the same group of people under similar conditions at different points in time. According to Kitchenham and Pfleeger [24], if the correlation between both of the answers is greater than 0.7 the test-retest reliability can be considered good.
- *Inter-Observer Reliability.* Ensures that the data collected in surveys or studies are consistent across different observers, enhancing the credibility and generalizability of the results. This is particularly important to mitigate interview bias in not self-administered surveys and to mitigate issues in qualitative analyses involving analyzing open-ended questions. It is typically addressed by having two or more observers involved in the interview and analysis process. For the analysis, it is common to measure the agreement between these raters, for instance, using Cohen's Kappa (for two raters) or Fleiss' Kappa (for three or more raters). Actions to improve the analysis consistency may involve having all raters assess a few cases and then discuss their ratings, which can help identify discrepancies and refine the rating process.
- *Statistical Generalizability.* Concerns ensuring that the results of a survey can be applied to a broader population beyond the specific sample studied. This is directly related to the sample representativeness and size, as well as to the inferential statistics applied to draw conclusions. This is a good point to remind the students of the importance of using robust sampling strategies and accurate inferential statistics. The sample's representativeness should be analyzed by comparing it against established benchmarks (when those exist). If the sample is representative, then inferential statistics can be used to draw conclusions (*e.g.*, testing hypotheses or calculating confidence intervals).

### 3.5.3 Assignments

A practical student assignment related to threats to validity and reliability could involve a critical analysis of existing published surveys. Students select and review specific surveys from software engineering literature, identify potential threats to both validity and reliability and discuss how these threats could affect the conclusions of the survey. Furthermore, they could propose and discuss possible mitigation actions for the identified threats.

> **Key Takeaways on Teaching Threats to Validity and Reliability**
>
> - Teach the concepts of validity and reliability with special attention. These are central to understanding the thoroughness and trustworthiness of software engineering surveys.
> - Teach and discuss the different types of validity relevant to survey research in software engineering.
> - Teach and discuss reliability concerns that are relevant to survey research in software engineering.

## 3.6 Teaching Ethical Considerations

Ethical considerations are paramount in survey research within software engineering, as they ensure ethical involvement of participants and ethical treatment of their data, the integrity of the data, and the credibility of the research findings. Educating students about these ethical concerns is essential for fostering responsible research practices. We need to emphasize the importance of considering thoughtfully how and whom we contact for a survey study [46].

While ethical issues have been discussed in empirical software engineering [36] and while there are well established regulations that touch upon selected aspects of ethics (such as privacy preservation by General Data Protection Regulatio - GPPR), there are still currently no established standards or guidelines on conducting surveys ethically in software engineering. However, general principles for survey research ethics from other areas [15, 33] can be directly applied to our context. These include following informed consent rules and respecting confidentiality and privacy.

As for the informed consent, participants must be fully informed about the nature of the research, what it involves, the risks and benefits, and their rights to withdraw at any time without penalty. With respect to privacy and confidentiality, researchers must protect the privacy of participants and the confidentiality of their data, using data encryption and anonymization techniques where appropriate. Students should also be taught how to report ethics considerations in software engineering publications [3].

Another important aspect is teaching students the importance of submitting survey research to institutional ethics review boards, as they will ensure the research adheres to ethical standards and protects participants. This process also educates students on the complexities of ethical research. In professional settings, corporate ethics committees perform a comparable role.

As shown in Table 2, an assignment for this lecture could involve students drafting an informed consent form for a software engineering survey based on what should typically be included in such document [3] and discussing ethical concerns related to risks and benefits, and potential implications.

---

**Key Takeaways on Teaching Ethical Considerations**

- Emphasize that ethics needs to be considered before contacting potential survey participants.
- Explain that participants must be fully informed about the nature of the research, what it involves, the risks and benefits, and their rights to withdraw at any time without penalty.
- Explain the role of the institutional ethics review boards and how to report survey ethics in software engineering publications.

---

## 4 Concluding Remarks

Throughout this chapter, we have explored effective strategies for teaching survey research, combining theoretical foundations with practical applications. From detailing a course syllabus to offering actionable advice based on seasoned experiences, the presented guidance aims to support educators in fostering a thorough understanding of survey research among their students. Additionally, we provide online resources, including slides for an entire course dedicated to teaching survey research.

We hope that the advice and materials provided in this chapter will help educators in setting up own courses or reflect upon their already existing courses when teaching survey research in software engineering. We emphasize that teaching material and advice will never be final and that educators are encouraged to adapt these materials as they see fit, customizing them to better suit their unique teaching styles and meet the specific needs of their students.

**Online Material**

In our online material repository https://zenodo.org/doi/10.5281/zenodo.11544897, the reader can find different slide decks that introduce the topic area of Teaching Survey Research in different educational settings and including units of different lengths.